\documentclass[journal=jacsat,manuscript=article]{achemso}
\pdfoutput=1
\usepackage[numbers]{natbib}
\usepackage{chemformula} 
\usepackage[T1]{fontenc} 
\usepackage{amssymb,amsmath,mathtools}
\usepackage[
colorlinks=true,        
allcolors = black,  
]{hyperref}
\hypersetup{
    colorlinks=true,
    linkcolor=black,
    filecolor=black,      
    urlcolor=black,
    pdftitle={Overleaf Example},
    pdfpagemode=FullScreen,
    }

\author{Arnab Barman Ray}
\affiliation{The Institute of Optics, University of Rochester, 480 Intercampus Dr, Rochester, NY 14627, USA}
\author{Trevor Ollis}
\affiliation{Department of Physics and Astronomy, University of Rochester, Rochester, NY 14627, USA}
\author{Sethuraj K. R.}
\affiliation{The Institute of Optics, University of Rochester, 480 Intercampus Dr, Rochester, NY 14627, USA}
\alsoaffiliation{Center for coherence and quantum optics, Department of Physics, University of Rochester, 480 Intercampus Dr, Rochester, NY 14627, USA}
\author{Anthony Nickolas Vamivakas}
\affiliation{The Institute of Optics, University of Rochester, 480 Intercampus Dr, Rochester, NY 14627, USA}
\alsoaffiliation{Center for coherence and quantum optics, Department of Physics, University of Rochester, 480 Intercampus Dr, Rochester, NY 14627, USA}
\alsoaffiliation{Materials Science, University of Rochester, Rochester, NY 14627, USA}
\alsoaffiliation{Department of Physics and Astronomy, University of Rochester, Rochester, NY 14627, USA}

\email{nick.vamivakas@rochester.edu, abarmanr@ur.rochester.edu}

\title[An \textsf{achemso} demo]
  {Diffusion of valley-coherent dark excitons in a large-angle incommensurate Moir\'{e} homobilayer}	

\abbreviations{IR,NMR,UV}

\begin{document}

\date{\today}

\begin{abstract}
The last few years have witnessed a surge in interest and research efforts in the field of twistronics, especially in low-angle twisted bilayers of transition metal dichalocogenides. These novel material platforms have been demonstrated to host periodic arrays of excitonic quantum emitters, interlayer excitons with long lifetimes, and exotic many-body states. While much remains to be known and understood about these heterostructures, the field of large-angle, incommensurate bilayers is even less explored. At twist angles larger than a few degrees, the presence of periodicity in these bilayers becomes chaotic, making the systems essentially aperiodic and incommensurate in nature due to the limitations of fabrication techniques. In this work, we demonstrate the emergence of a brightened dark intralayer exciton in twisted n-doped molybdenum diselenide homobilayer. We show that this dark exciton diffuses across the excitation spot more efficiently as compared to bright trions or excitons, reaching diffusion lengths greater than 4 microns. Temperature-dependent spectra provide corroborative evidence and reveal a brightened dark trion. Almost inexplicably, this dark exciton showcases a robust valley coherence, which we attribute to a small mixing of the spin-resolved conduction bands due to an absence of out-of-plane reflection symmetry arising from a strong dielectric contrast. Our results reveal some of the richness of the physics of these large-angle systems while uncovering new opportunities for valleytronic devices that may utilize these more valley-robust "mixed" dark excitons.

\end{abstract}
	
\maketitle


\section{Introduction}
Moir\'{e} hetero- and homo-bilayers of transition metal dichalcogenides(TMDCs) have been shown to host correlated electronic phenomena \cite{1,2,3,4,5,6,7,RN40} and arrays of programmable quantum emitters \cite{8,9,10,11,12,13,14, RN147, NME}. Less explored has been the physics of large-angle twisted homobilayers ($ 10^{\circ} < \theta < 50^{\circ})$. While the Moir\'{e} superlattice is active only at small angles \cite{15} and allows for the trapping of excitons, at high angles, this periodicity is broken. The condition for commensurability or periodicity in a twisted bilayer system of two honeycomb lattices is provided by the equation\cite{17}, $\cos{\theta} = \frac{3m^2+3mn+n^2/2}{3m^2+3mn+n^2}$, where $\theta$ is the twist angle and $m$ and $n$ are a pair of coprime positive integers. Immediately, it can be discerned that angles where the value of the cosine is irrational do not exhibit periodicity. However, given that there are an infinite number of coprime pairs of positive integers, it is possible to have a rational number expressed in the form of the previous equation arbitrarily close to the irrational number in question. This leads to a highly sensitive, if not wholly chaotic period at large angles\cite{16}. With the limitations of current fabrication techniques\cite{18}, this leads to an essentially aperiodic structure without the band modulating and flattening effects that a Moir\'{e} superlattice usually endows.

Without an active and periodic Moir\'{e} superlattice, these large-angle bilayers may seem to hold little promise in terms of the originally intended applications in quantum computation for these systems. A high twist angle pushes diffusive interlayer excitons\cite{RN39} out of the light cone while limiting the possibility of simulating correlated states in periodic lattices\cite{RN36} due to an effective uncoupling of electronic communication between the two layers, as is seen in graphene. However, as we show in this work, this class of aperiodic bilayers can be interesting in its own right. We focus on a large-angle twisted homobilayer of n-doped molybdenum diselenide. These systems have been explored in a recent work\cite{19} over a range of twist angles. We show, using standard micro-photoluminescence (PL) experiments, that at large twist angles, the proximity of a highly polarizable monolayer to the other alters its optical properties. As reported previously \cite{19}, we observe a large redshift of the trionic and excitonic resonances in the bilayer. This redshift can be partly attributed to the proximal monolayer's large polarizability, similar to the much smaller redshift experienced by a free monolayer when encapsulated by a high dielectric constant insulator hBN \cite{20}. Thus, the large-angle twist of the bilayer serves to prevent the heterostructure from becoming an indirect bandgap semiconductor which is the case for a Bernal-stacked ($0^{\circ}$) bilayer \cite{21}, while at the same time having the Moir\'{e} superlattice potential inactive.  In this regard, it is worth noting that layer-hybridized indirect excitons in natural homobilayers have also been investigated for diffusion control\cite{RN148}.

Upon investigating the spatial diffusion of bound complexes under steady-state CW excitation at low intensities across the PL spectrum, we uncover the quenching of diffusion lengths at energies of large quantum yield or maximum PL intensity. Investigating diffusion lengths points us toward the presence of a more diffusive bound species at a slightly higher emission energy than the bright spin-singlet exciton. By carefully deconstructing the spectra, we uncover that this new species of trion diffuses more efficiently than the bright excitons or trions. Investigating the temperature dependence of the PL reveals evidence for a population transfer to this brightened dark trion. Population transfer to higher-energy dark excitonic states is what is responsible for the decrease in quantum yield with increasing temperature in single monolayers of $\text{MoSe}_2$ and $\text{MoS}_2$ with the point group symmetry $D_{3h}$, and is well documented and understood\cite{22,23,24}. However, we find that for a large-angle twisted bilayer this population transfer still allows us to capture some of the PL emitted from this spin-forbidden intervalley dark trion, as a result of the removal of the out-of-plane reflection symmetry - causing the point group of both the two monolayers to reduce to $D_{3v}$. The emission due to these dark excitons is usually suppressed due to the vastly smaller radiative rate that accompanies spin-flip electronic transitions in monolayers\cite{RN131}. This brightening of this triplet trion is similar to what has been predicted and observed for the case of interlayer excitons in $\text{WSe}_2-\text{MoSe}_2$ heterobilayers\cite{25,26}. Furthermore, we show that these dark triplet trions are more diffusive than their bright counterparts, with diffusion lengths exceeding 4 microns. Dark excitons/trions are usually long-lived, optically decoupled from the environment, and serve as a reservoir for their bright counterparts, playing a crucial role in the condensation of excitons in other well-studied semiconductor platforms\cite{27,28}. Hence it is important to understand their properties.

\section{Results and Discussion}

\begin{figure}[h]
\centering
\vspace{0.5cm}
\includegraphics[height=0.3\textwidth,width=0.9\textwidth]{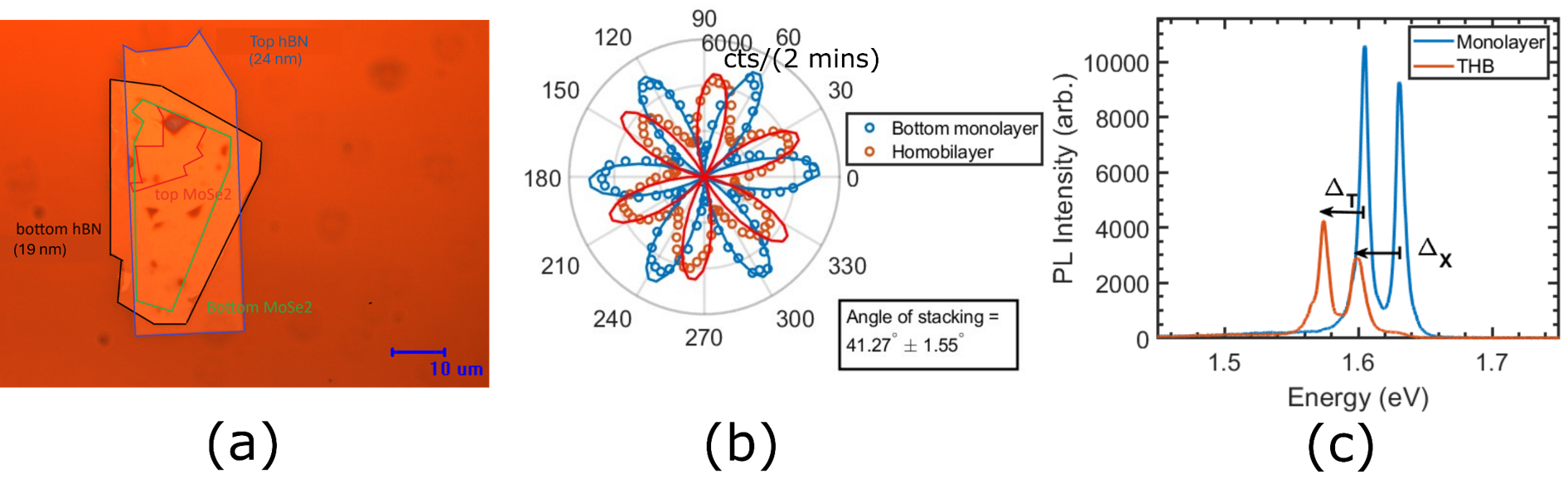}
\caption{ (a) Optical micrograph of the homobilayer,(b) co-polarized second harmonic signal from the bottom monolayer and the bilayer used to estimate the twist angle, (d) PL spectrum of the monolayer and bilayer with the shifted trionic (T) and excitonic (X) peaks at 4K, 532 nm CW excitation.}
\label{Fig1}
\end{figure}

Fig.~\ref{Fig1}(a) provides an optical micrograph of the homobilayer investigated in this work. We mechanically exfoliate and assemble monolayers of n-type $\text{MoSe}_2$ (2D semiconductors) and thin, flat flakes of hBN (2D semiconductors) on a distributed Bragg reflector (see Methods) with its reflection band centered at 770 nm to optimize the PL signal collection. Using a pulsed ti-sapphire laser at 790.5 nm, we collect the co-polarized second harmonic signal generated from the more accessible bottom monolayer and the homobilayer as a function of the laser polarization angle. After accounting for the effects of the beamsplitters in the signal and collection path, the corrected SHG signal is presented in Fig.~\ref{Fig1}(b) with their respective fits, revealing the twist angle to be: $41.27^{\circ} \pm 1.55^{\circ}$. We note that the SHG signal collected also validates the high quality of the fabricated sample and that there is minimal strain present\cite{29,30} away from the visible bubbles in the micrograph. We probe the photoluminescence spectra in a confocal microscopy setup (0.70 NA microscope objective) where the sample is cooled to cryogenic temperatures (all measurements are at 6 K unless specified). Fig.~\ref{Fig1}(c) shows the PL signal from the monolayer and bilayer. We observe a large redshift of the trionic and excitonic resonances of $\Delta_T = 30.4 \text{ meV}$ and  $\Delta_X = 32.8 \text{ meV}$, due to enhanced dielectric screening.

Next, we shift our attention to the diffusion of the excitonic and trionic complexes across the emission spot in both the monolayer and the bilayer under steady-state excitation. While it has been demonstrated that at small angles the localizing effects of the Moir\'{e} potential impedes the diffusion for interlayer and intralayer excitons\cite{31,32, RN37}, the case for large-angle bilayers is less explored either theoretically or experimentally. We focus on the diffusion lengths obtained for different species in the monolayer and bilayer. We note that further experimental work involving measurements of PL lifetime would help calculate the diffusion coefficients. However, the focus of this work is to chronicle how diffusion lengths were used to identify the brightened spin-forbidden dark exciton.

\begin{figure}[h]
\centering
\vspace{0.1cm}
\includegraphics[height=0.65\textwidth,width=1.00\textwidth]{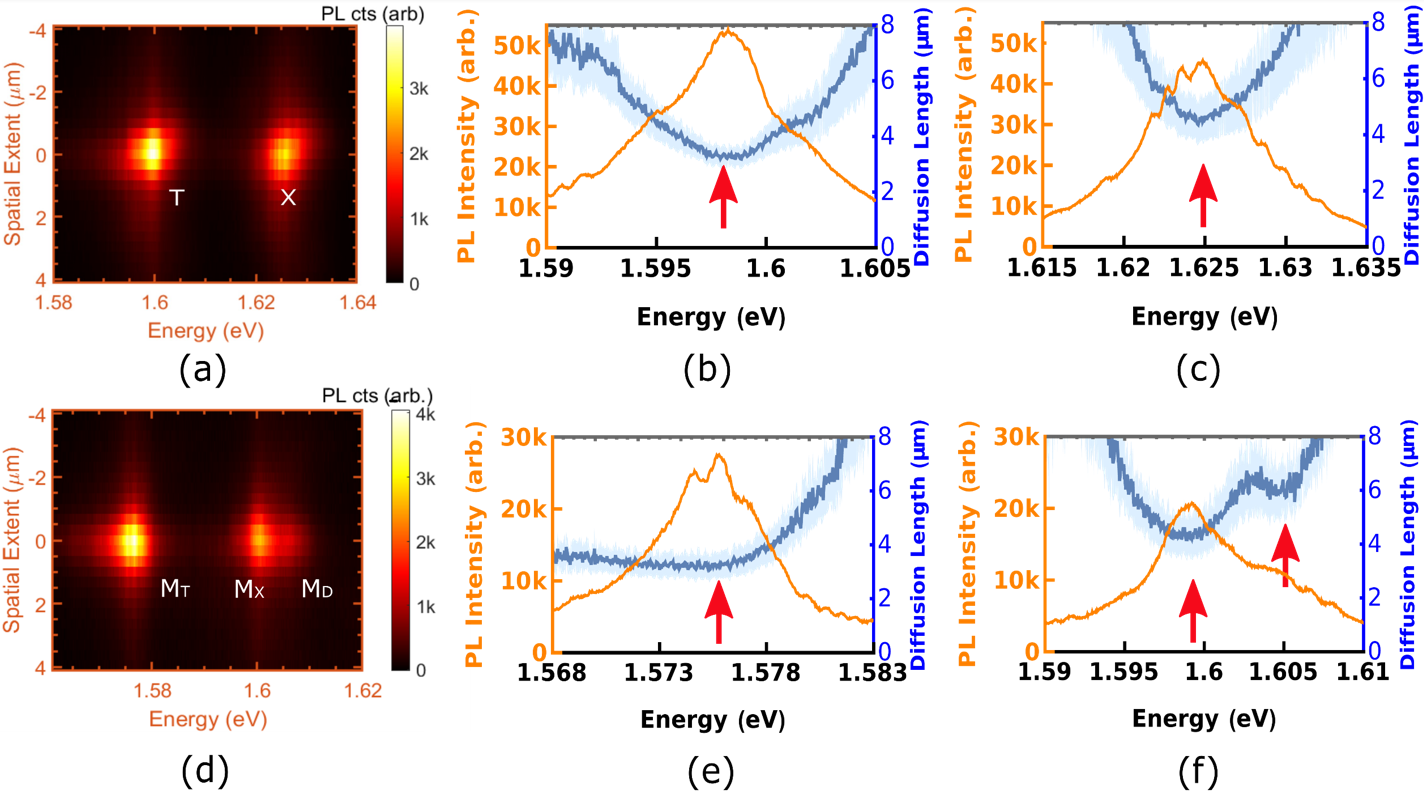}
\caption{ (a) Spatio-spectrum of the monolayer, PL spectrum and the spectral variation of diffusion lengths of (b) the trion (T), and (c) the exciton (X). (d) Spatio-spectrum of the homobilayer and the spectral variation of diffusion lengths of (b) the trion ($\text{M}_{\text{T}}$), and (c) the exciton ($\text{M}_{\text{X}}$). Deep blue lines trace the obtained diffusion lengths while light blue shaded regions demarcate the 95 \% confidence intervals from the fits.}
\label{Fig2}
\end{figure}

Upon excitation with a power of $10\, \mu W$ in a spot of radius $ \sim 1 \,\mu m$, we image the emission spot on the CCD camera of our spectrometer setup and record the spatio-spectrum of the monolayer and the bilayer in Fig.~\ref{Fig2}(a) and (d). Under steady-state excitation, neglecting the effect of exciton-exciton interactions, we fit the spatial extent of the PL intensity as a function of emission energy using the diffusion equation, $n(x) \propto \int_{-\infty}^{\infty} dx^\prime K_0 \left(\frac{x^\prime}{L_D}\right)e^{-\frac{(x-x^\prime)^2}{L^2}}$, where $K_0$ is the modified Bessel function of the second kind\cite{31,33,34}. We determine the laser spot linewidth $L$ by fitting it with a Gaussian function. This fit provides us with the diffusion lengths $L_D$ as a function of spectra. We verify that the diffusion lengths thus obtained do not change considerably over three orders of magnitudes of the excitation intensity in the Supporting Information (see Fig. S1). Note that the diffusion equation is less appropriate for modeling trion diffusion due to local electric field effects from donor atoms which can modify their dynamics \cite{35}. However, the quenching of the diffusion lengths for trions at energies corresponding to high quantum yield (high PL intensity) denoted by red arrows in Fig.~\ref{Fig2}(b), (c), (e) and (f) highlights the accuracy of our measurements. Moreover, our data captures the fact that trions diffuse less as compared to excitons due to their larger effective mass\cite{36} and the aforementioned effects which is well documented in the literature\cite{35,37}. Fig.s ~\ref{Fig2}(b), (c), (e) and (f) show that the spectral variation of the diffusion lengths qualitatively resembles a reflection of the PL spectra about the horizontal axis. 

For any diffusive bound complex, the rate of population decay is given as a sum of the radiative and non-radiative rates, $\Gamma_{tot} = \Gamma_{r} + \Gamma_{nr}$. The PL lifetime is given as $\tau_{tot} = \frac{\tau_{r}\tau_{nr}}{\tau_r+\tau_{nr}}$, where $\tau_{r} = \Gamma_{r}^{-1}$ and $\tau_{nr} = \Gamma_{nr}^{-1}$. As the diffusion lengths theoretically are given by $L_{D} = \sqrt{D\,\tau_{tot}}$ (where $D$ is the diffusion coefficient), substituting this gives, $L_{D} = \sqrt{D \frac{\tau_{r}\tau_{nr}}{\tau_r+\tau_{nr}}}$. Using the relation satisfied by the intrinsic PL quantum yield given as $\text{QY} = \frac{\Gamma_r}{\Gamma_r + \Gamma_{nr}}$, it is straightforward to arrive at the equation, $L_D = \sqrt{D\,\tau_{nr}(1-\text{QY})}$. This relation explains the quenching of the diffusion lengths at energies of high PL intensity or quantum yield.

\begin{figure}[h]
\centering
\vspace{0.1cm}
\includegraphics[height=0.70\textwidth,width=1.00\textwidth]{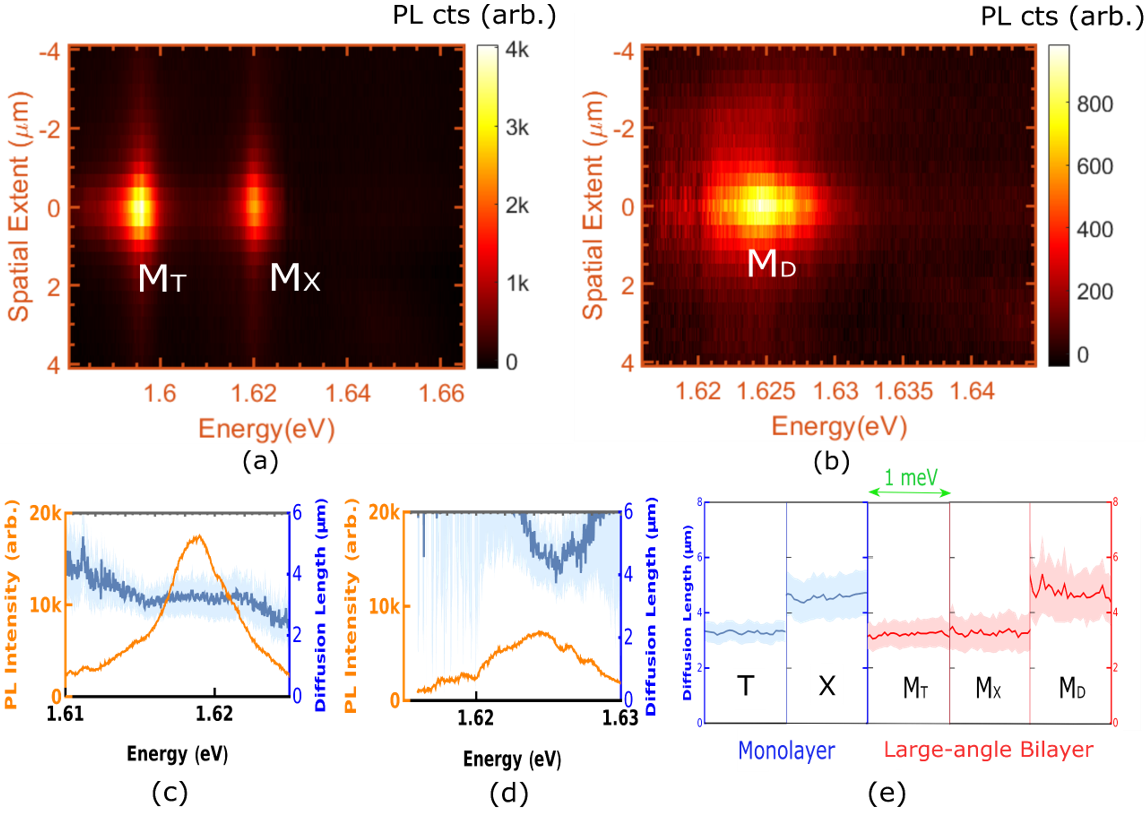}
\caption{ (a) Corrected spatio-spectrum of the bilayer trion and bright exciton and (b) the spectral variation of diffusion length for the bright exciton. (c) Corrected spatio-spectrum of the bilayer dark exciton and (d) the spectral variation of diffusion length for the dark exciton. Deep blue lines trace the obtained diffusion lengths while light blue shaded regions demarcate the 95 \% confidence intervals from the fits. (e) Diffusion lengths of different species in the monolayer and bilayer over a spectral width of 1 meV across their respective peak PL intensities. Deep blue (red) lines trace the obtained diffusion lengths while light blue(red) shaded regions demarcate the 95 \% confidence intervals from the fits for the monolayer (bilayer).}
\label{Fig3}
\end{figure}

Fig. ~\ref{Fig2}(d) and (f) hints at the presence of a less bright species of exciton (which we label $\text{M}_\text{D}$) at a slightly higher energy than the bright exciton $\text{M}_\text{X}$. At this point, it is impossible to ascertain whether this species is fundamentally different from the bright exciton or whether it is a result of inhomogeneity or dielectric disorder\cite{37} in the sample. Moreover, while Fig.~\ref{Fig2}(e) and (f) seem to indicate that both $\text{M}_\text{X}$ and $\text{M}_\text{D}$ diffuse much more efficiently than compared to $\text{M}_\text{T}$ or even the monolayer exciton X, the spectral proximity of these two species may cause leakage of the tails of their respective spectrums at the peak energies of each other, and thus artificially inflating their actual diffusion lengths. To circumvent this problem, we try to disentangle the contributions of each of these two species. We find that the spectral slices that make up the spatio-spectrums lend themselves well to double-lorentzian fits (see supporting Fig. S2). We are thus able to investigate the diffusion lengths of both of these species separately in Fig.~\ref{Fig3}.

\begin{figure}[h]
\centering
\vspace{0.1cm}
\includegraphics[height=0.70\textwidth,width=1.00\textwidth]{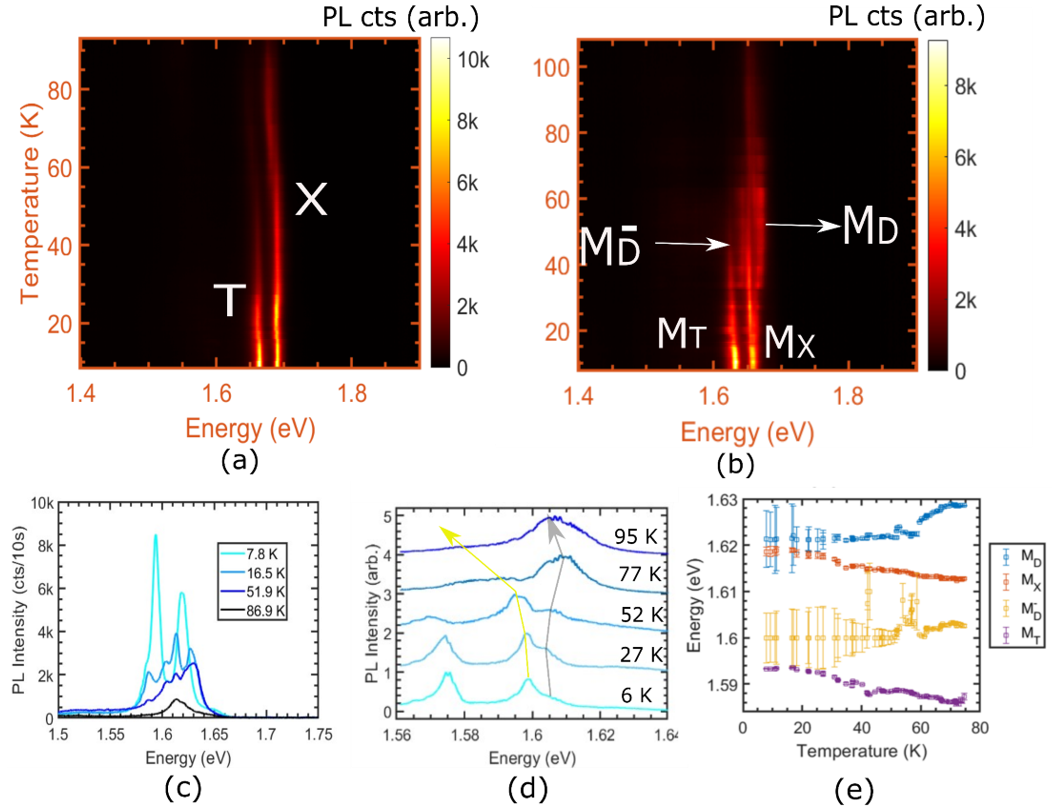}
\caption{ Evolution of PL with temperature for the (a) monolayer and (b) bilayer. (c) PL spectra at different temperatures exhibitng four separate peaks. (d) Dependence of peak energies of bright exciton (yellow arrow) and dark exciton (dark arrow) with temperature. (e) Extracted peak energies as a function of temperature.  }
\label{Fig4}
\end{figure}

We next compare and contrast the diffusion lengths of different species in Fig.~\ref{Fig4}. For the monolayer, the exciton diffuses more efficiently than the trions, which face an inward electrostatic force from donor atoms, altering their diffusion significantly. For the bilayer, our results indicate that despite the absence of a periodic Moir\'{e} superlattice, the diffusion of bright excitons is suppressed. We suggest that the suppression of excitonic diffusion in these bilayers may arise from induced dipole interactions between the bright excitons and the donor atoms across both monolayers, leading to a qualitatively different behavior than the bright excitons in monolayers. We note that the less bright excitonic species $\text{M}_{\text{D}}$ diffuses more efficiently as bright excitons in the monolayer, which may arise from a relatively longer lifetime\cite{39,40}.

\begin{figure}[h]
\centering
\vspace{0.1cm}
\includegraphics[height=0.5\textwidth,width=1.00\textwidth]{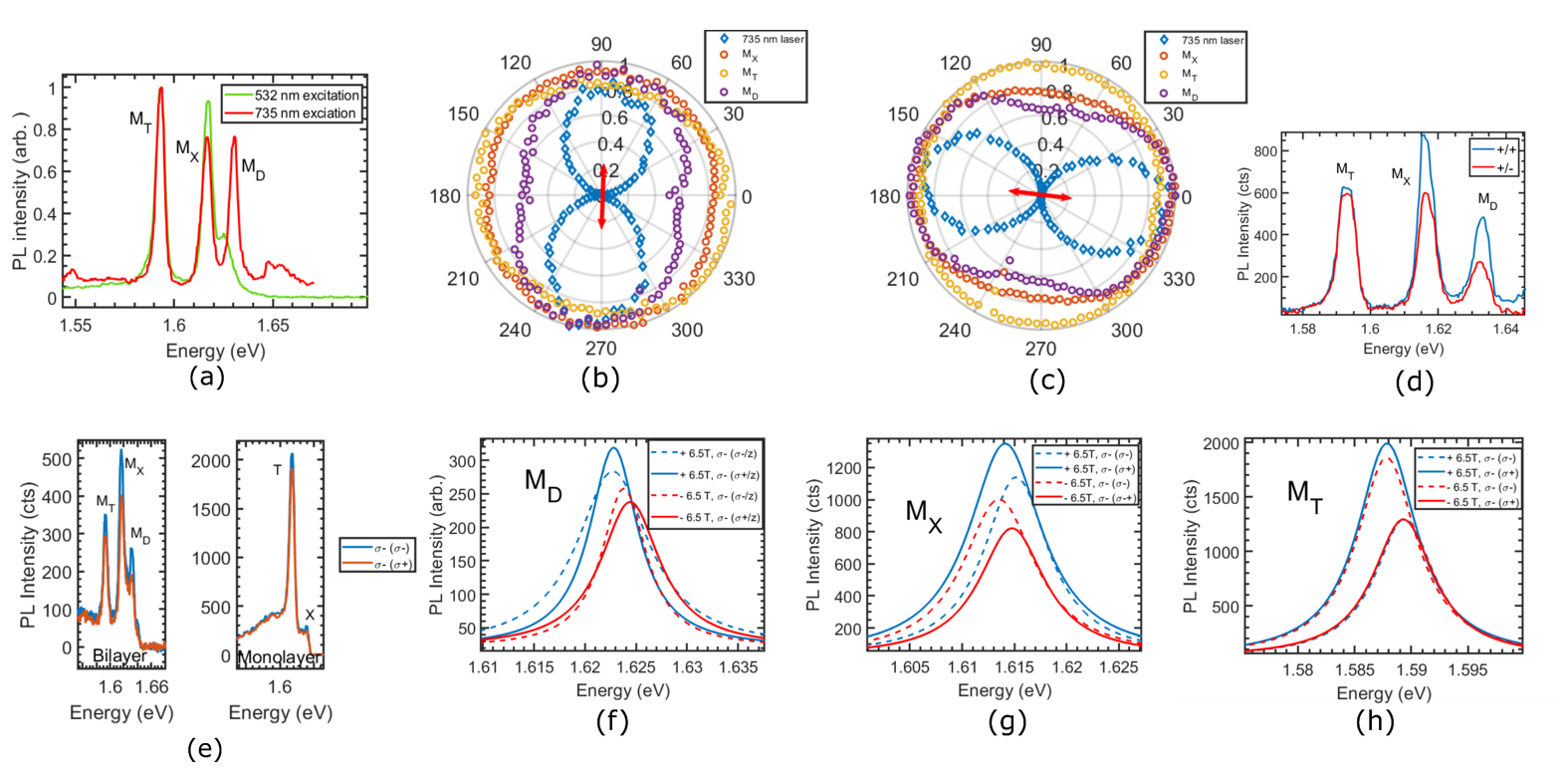}
\caption{(a) PL spectrum of the bilayer at different excitation photon energies, (b) and (c) integrated PL intensities of the three species (from Lorentizan fits) and the laser, as a function of detection polarization angle at an excitation power of 28 $\mu$W, (d) PL spectrum of bilayer for co- and cross-polarizations with respect to the linearly polarized laser, (e) net valley polarization of the bilayer (left) and monolayer (right) with 735 nm excitation at zero magnetic field, (f),(g) and (h) separate fits of PL spectra for the three species(dark exciton, bright exciton and bright trion respectively) with circular polarization selection and magnetic field, with 532 nm excitation at 50 $\mu$W. Sample temperature was kept at 12.5 K for all these measurements.}
\label{Fig5}
\end{figure}

To determine the nature of the more diffusive species, we trace the PL signal as a function of the sample temperature. The evolution of diffusion lengths with temperature is provided in the Supporting Information. Fig.~\ref{Fig5}(a) exhibits the temperature dependence of the PL from the monolayer. We notice the monotonic redshift with increasing temperature\cite{41} and the decrease in PL yield. This is due to the presence of higher energy dark states in $\text{MoSe}_2$. This decrease in quantum yield is opposite to that of tungsten-based TMDC monolayers, where the presence of low-lying dark states leads to an increase in PL yield with increasing temperature. We trace the PL from the bilayer in Fig.~\ref{Fig4}(b). We note evidence for a visible population transfer to the now-brightened dark states at around $30\,\text{K}$, which corresponds to a thermal energy of $\sim 2.5\,\text{meV}$, about half of the difference in the peak energies of $\text{M}_{\text{X}}$ and $\text{M}_{\text{D}}$. Around that temperature, we detect evidence of a brightened dark trionic state $\text{M}_{\text{D}}^{-}$\cite{42} (see Supporting Fig. 5). The extra binding energy of the dark trion at $30\,\text{K}$ is $24\,\text{meV}$ and is close to the binding energy of the bright trion ($27\,\text{meV}$) at the same temperature. The four excitonic and trionic species in question are clearly identifiable in the spectra as four separate peaks at $16.5\,\text{K}$ in Fig.~\ref{Fig4}(c). The dark species investigated in this work are intravalley direct dark excitons and trions, and the PL emission is not phonon-assisted, which can be surmised from the relative positions of their peak energies from that of their bright counterparts. Finally, we report the unusual, non-monotonic behavior of the energy of the dark exciton in Fig.~\ref{Fig4}(d) and (e). In contrast to the continuous redshift of the bright exciton with temperature, the dark exciton initially undergoes a considerable blue shift in its energy before it starts to redshift. This, too, points at a different band origin of the electron in the dark exciton. The cause for this behavior may be analogous to the anti-funneling effects observed for momentum-forbidden dark excitons\cite{43} in single tungsten-based monolayers, which also arises from a difference in how the electron bands evolve under strain, or in this case, temperature.

Finally, we study the polarization-resolved properties of the observed bound complexes in a second sample ($40^{\circ} \pm 2^{\circ}$, tear-and-stack) with a high-NA (0.82) confocal microscope. We find that using a near-resonant laser energy leads to a better-resolved PL spectra and preferential formation of the dark exciton in Fig.~\ref{Fig5}(a). By investigating the quantum valley coherence of the three species, we found, to our surprise, that the dark exciton exhibits an improved and robust valley coherence as opposed to the other excitonic resonances in Fig.~\ref{Fig5}(b), (c) and (d). The species also demonstrate an appreciable amount of valley polarization with 735 nm excitation with no magnetic field (Fig.~\ref{Fig5}(e)). The valley coherence substantiates the claim that the $\text{M}_{\text{D}}$ emission does not arise from a disorder/defect as these emitters are usually linearly polarized and do not follow the excitation laser polarization\cite{RN145}. 

The data are surprising for two reasons - the first being that $\text{MoSe}_2$ monolayer is exceptional among its family of TMDCs in that excitons, while bright at cryogenic temperature, do not possess any appreciable valley polarization (or coherence) with nonresonant or near-resonant excitation\cite{RN138}. Several reasons have been suggested for this in the literature, ranging from D'yakanov-Perel', Elliott-Yafet and MSS mechansims \cite{RN138, RN142, RN139} as well as a resonance of an optical phonon mode with the conduction-band spin splitting \cite{RN144}. Secondly, the emission from a dark exciton in $\text{MoSe}_2$ monolayer is originally z-polarized. While collection by a high-NA infinity-corrected objective is possible, this should convert the z-polarization to a radially polarized beam\cite{RN146} which should be insensitive to selection by a quarter-wave plate/linear-polarizer combination, especially to a bucket detector such as our fiber-spectrometer-CCD combination. Hence, we investigate the valley splitting of the species with a magnetic field in a Faraday configuration. We find, in Fig.~\ref{Fig5}(f), (g), and (h), that in contrast to the bright species, the dark exciton displays an equivalent energy shift for emission of both handedness under excitation of a single valley with circularly polarized light. This indicates that the majority of the PL emission is from a single valley, while indicating that a substantial portion of the PL collected is primarily z-polarized (radially polarized). Furthermore, the presence of a non-zero valley polarization and the subsequent valley coherence indicates that the emission from dark exciton is not purely z-polarized and hints at a spin-mixing of the two conduction bands due to the broken symmetry of the bilayer. These "mixed" dark excitons seem to be comparatively well-shielded from the intervalley scattering processes that plague the bright excitons in both the bilayer and the monolayer.

\section{Conclusion}
To summarize, we uncover the brightening of the spin-forbidden dark exciton and dark trion in a large-angle incommensurate Moir\'{e} homobilayer. We identify a more diffusive species by analyzing the spectral variation of diffusion lengths in the PL spectrum, which we assign to the dark exciton. Investigating the temperature dependence of the PL spectrum leads us to discover the population transfer effects that are otherwise undetectable in the monolayer. We uncover that these dark excitons are slightly mixed, and display a degree of valley addressability which is more robust that its bright counterparts. Diffusive, robust, valley-addressable dark excitons may pave the way to future valleytronic devices. Our results uncover several interesting facets of exciton photophysics is these less-explored large-angle bilayer systems.

\section{Methods}

The first author prepared the manuscript, and all the co-authors participated in discussing the draft. After the draft was near its final stage, ChatGPT 4o was used to identify grammatical errors and consistency issues and to improve the general readability of the text. The prompt and the response from ChatGPT 4o are provided in the supplementary information.

\subsection{Fabrication}
The monolayers and hBN (high-pressure anvil growth) were mechanically exfoliated from high-quality bulk samples obtained from 2D Semiconductors. The indivdual flakes were then assembled step-by-step under an optical microscope using dome-shaped windows constructed from cured PDMS with a thin pane of PPC (poly-propylene carbonate). After the device was constructed the assembly was heated to release it when in contact with the DBR chip (with an additional 98 nm of $\text{SiO}_2$ on top). The SiN-terminated distributed Bragg reflector was fabricated using a PECVD method, where 10.5 pairs of SiN/$\text{SiO}_2$. the second sample was fabricated similarly, with a tear-and-stack technique on an $\text{SiO}_2$/Si substrate.

\subsection{Measurements}
The measurements were carried out using a custom-made confocal microscope. A 532 nm DPSS laser is focused into a submicrometer diameter spot using a 0.70 NA objective lens in a closed-cycle cryostat (Montana Instruments) at 6 K. The emission spot is relayed through the objective and imaged onto the CCD camera using an achromatic lens system. One of the achromatic lenses is stepped longitudinally to minimize the effects of longitudinal color in the system between the measurements of the laser spot and the photoluminescence as they differ in their wavelengths considerably. The collected PL is then analyzed using a Princeton Instruments spectrometer (Acton SP-2750i) and an LN2 cooled Pylon CCD camera. An Msquared Sprite XT femtosecond pulsed ti-saph (80 MHz repetition rate) is used at 790.5 nm for the second harmonic generation measurements. A Coherent Chamaeleon II laser is used its Continuous-Wave operation (alignment) mode at 735 nm for some of the measurements. Measurements on the second sample were performed in an Attodry 1000 magneto-optics setup with a 0.82 NA objective and a home-built polarization sensitive microscope. 

\section{Supporting Information}
More data and analysis complementing the results presented in the main paper are presented in the supporting information file.

\section{Acknowledgments}
	
This work was supported by
AFOSR FA9550-19-1-0074 from the Cornell Center for Materials Research. S.K.R. acknowledges support from AFOSR FA9550-21-1-0322 and DARPA YFA D19AP00042.


\bibliography{biblio2}

\end{document}


\subsection{Excitation intensity dependence of diffusion lengths}

\begin{figure}[h]
\centering
\vspace{0.5cm}
\includegraphics[height=0.80\textwidth,width=1.00\textwidth]{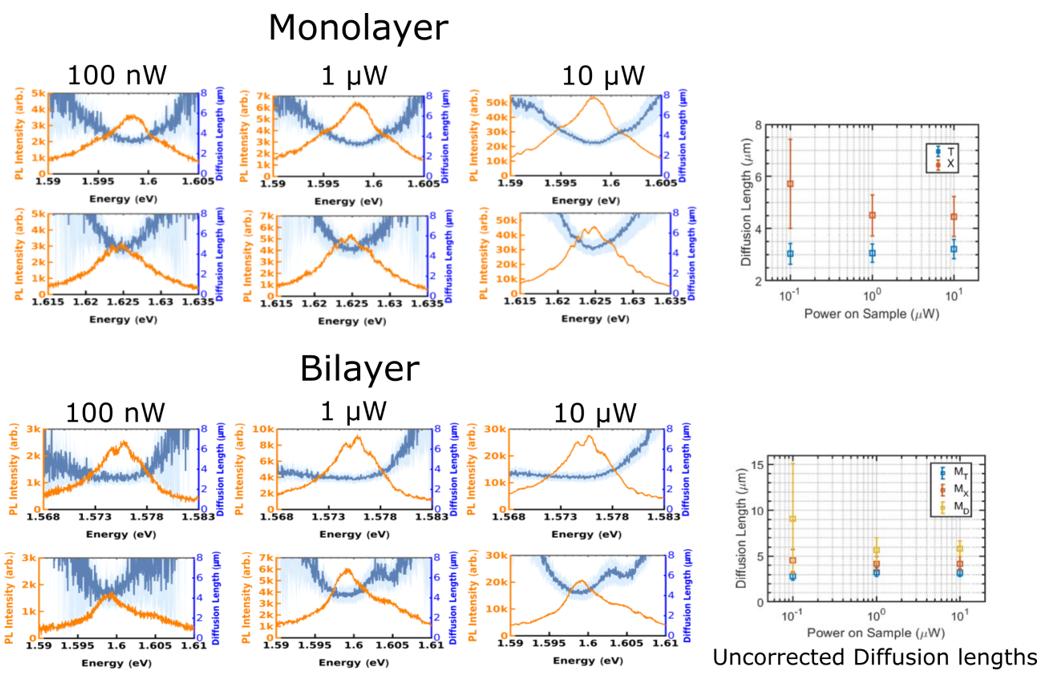}
\caption{Diffusion lengths and spectra for the monolayer and bilayer over three order of magnitudes of excitation power, showcasing the validity of the simplifying assumptions in the paper.}
\label{fig1}
\end{figure}

Diffusion of a gas of dipolar excitons is governed by the equation, $\frac{\partial \rho(\vec{r},t)}{\partial t} = D \nabla^2 \rho(\vec{r},t) + \mu I_0 \nabla{(\rho(\vec{r},t)\nabla\rho(\vec{r},t))} - \frac{\rho(\vec{r},t)}{\tau_{tot}} + g(\vec{r},t)$, where $\rho(\vec{r},t)$ is the density of excitons, $\mu$ is the mobility, $I_0$ is the pair interaction energy and $g$ is the generation rate (other symbols have their usual meanings from the main text)\cite{1,2,3}. This equation remains valid for the non-dipolar (without a permanent dipole moment) intralayer excitons considered in this work, especially at high densities\cite{4}. However, at the low densities considered here and under the assumptions of steady state, the equation is simplified to retain only the effects of thermodynamic diffusion, generation, and decay to give, $D \nabla^2 \rho(\vec{r},t) = \frac{\rho(\vec{r},t)}{\tau_{tot}} - \alpha I \int\int dA \exp{(\frac{-r^2}{2\sigma^2})}$, where $\alpha$ is the absorption coefficient and $I$ is the laser intensity, considering a gaussian profile for the laser spot. This equation can then be solved to provide the form for the steady-state profile of excitons discussed in the main text.

To substantiate the validity of the above assumptions, the uncorrected ($\text{M}_{\text{X}}$ and $\text{M}_{\text{D}}$ are spectrally mixed and affect each others' values of diffusion lengths) diffusion lengths are extracted over three orders of magnitudes of the excitation power in supporting figure.~\ref{fig1} for both the monolayer and the bilayer. The diffusion lengths are reported at the sites of quenching. As we can see at 100 nW of excitation power, the noise in the spectrum leads to large errors in the diffusion lengths for some of the species. However, despite this, we do not observe any measurable qualitative difference in the diffusion lengths for any of the species, within the bounds of error, over the ranges of excitation power considered here. 

\subsection{Disentangling contributions of bright and dark excitons}

\begin{figure}[h]
\centering
\vspace{0.5cm}
\includegraphics[height=0.35\textwidth,width=0.80\textwidth]{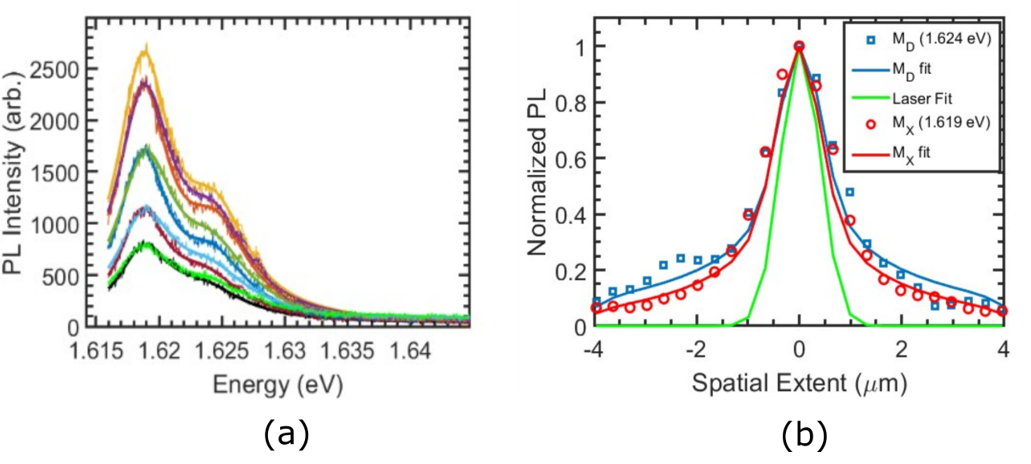}
\caption{(a) Several spatio-spectral slices and their double-lorentzian fits, (b) Corrected spatial profile of bright and dark excitons and their respective fits, along with the fit used for the 532 nm laser spot.}
\label{fig2}
\end{figure}

In the bilayer, the spectral proximity of the bright and dark excitons causes the photoluminescence from one of the species to affect the apparent diffusion length obtained at the maximum PL intensity of the other. To remedy this we subject the spatio-spectral slices to fits of the sum of two lorentzian functions. We find that the slices lend themselves well to such a fit in supporting figure.~\ref{fig2} (a), which illustrates this for some representative slices. We can then subtract the contributions of the dark exciton from the spectra of the bright exciton and vice-versa. Supporting figure.~\ref{fig2} (b) showcases the spatial profile of the two excitons alongwith their fits as well as the gaussian fit for the laser spot used in the experiments. One of the achromatic lenses used to image the emission spot at the entrance slit of the spectrometer-CCD setup was stepped between imaging the laser spot and the PL emission to ensure minimal defocus at both wavelengths. The small asymmetry in the diffusion of dark excitons may arise from an interplay between dark-gray exciton interconversion as well as from a more sensitive(to dielectric disorder) dipole radiation profile of the gray excitonic contribution. This fine-structure of the dark exciton is too small to be resolved in this work\cite{5}.

\subsection{Second harmonic generation measurements}

\begin{figure}[h]
\centering
\vspace{0.5cm}
\includegraphics[height=0.35\textwidth,width=0.80\textwidth]{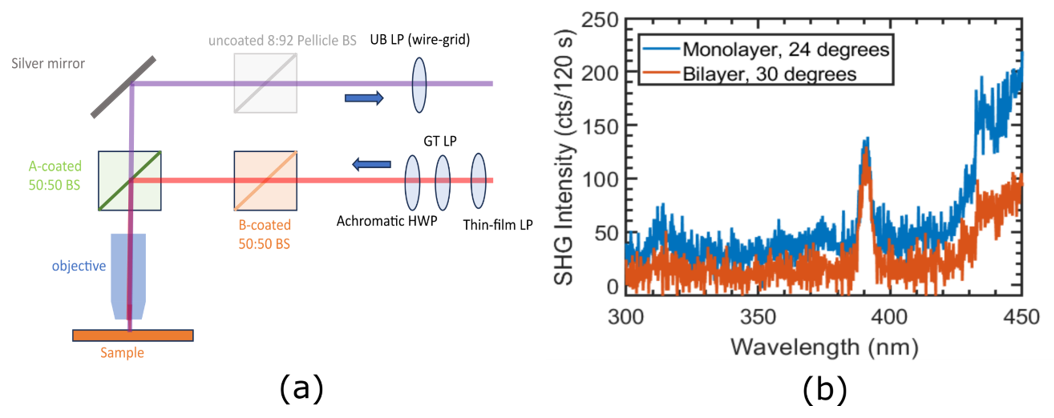}
\caption{(a) Setup used for the SHG measurements. (b) An example of the co-polarized SHG spectra from the monolayer and the bilayer at different laser polarization angles}
\label{fig3}
\end{figure}

Two Thorlabs Elliptec piezo-rotators(ELL14) with internal coordinate references, were used synchronously to measure the co-polarized second harmonic generation signal from the monolayer and the bilayer. A Thorlabs Achromatic half-wave plate(AQWP10M-980) was used to rotate the polarization of the incident femtosecond pulsed laser at 790.5 nm. The effects of the beamsplitters in the path of the incident laser and the outgoing signal were accounted for using reflection and transmission curves obtained from Thorlabs. The effect of the beamsplitters on the linear polarization direction of either the laser or the signal was calculated to be minimal.

\subsection{Temperature dependent measurements}

\begin{figure}[h]
\centering
\vspace{0.5cm}
\includegraphics[height=0.50\textwidth,width=1.00\textwidth]{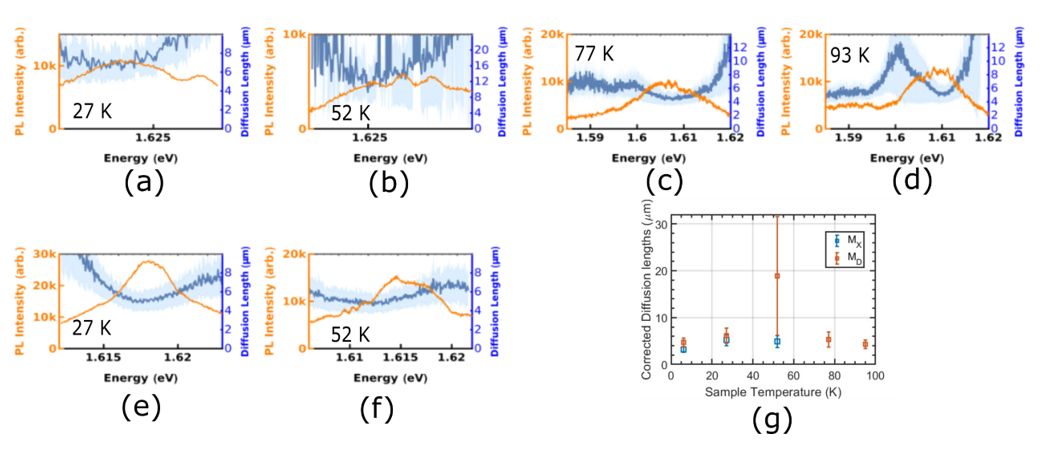}
\caption{(a)-(d) Corrected dark exciton spectrum and diffusion lengths at the indicated temperatures, (e)-(f) corrected bright excitonic spectrum at indicated temperatures, (f) extracted diffusion lengths at maximum PL intensities. All data refers to the bilayer. }
\label{fig4}
\end{figure}

We have investigated the temperature dependence of the diffusion lengths of the bilayer in this work. We see that between 6 K and 52 K, there is a noticeable increase in the diffusion lengths of both the bright and dark excitons. This can be attributed to an increased phonon interaction at higher temperatures\cite{6}. However, the very same mechanism is responsible for a broadening of the PL response. This together with the overall decrease in PL yield and the limitations in modelling the emission profile with lorentzian functions results in a large error in the diffusion length for the dark exciton at 52 K. Beyond 52 K, the negligible contribution from the bright excitons prevents us from obtaining an accurate estimate of the diffusion length. However, we do report a decrease in the diffusion length of the dark exciton over this range of temperature.

\begin{figure}[h]
\centering
\vspace{0.5cm}
\includegraphics[height=0.50\textwidth,width=1.00\textwidth]{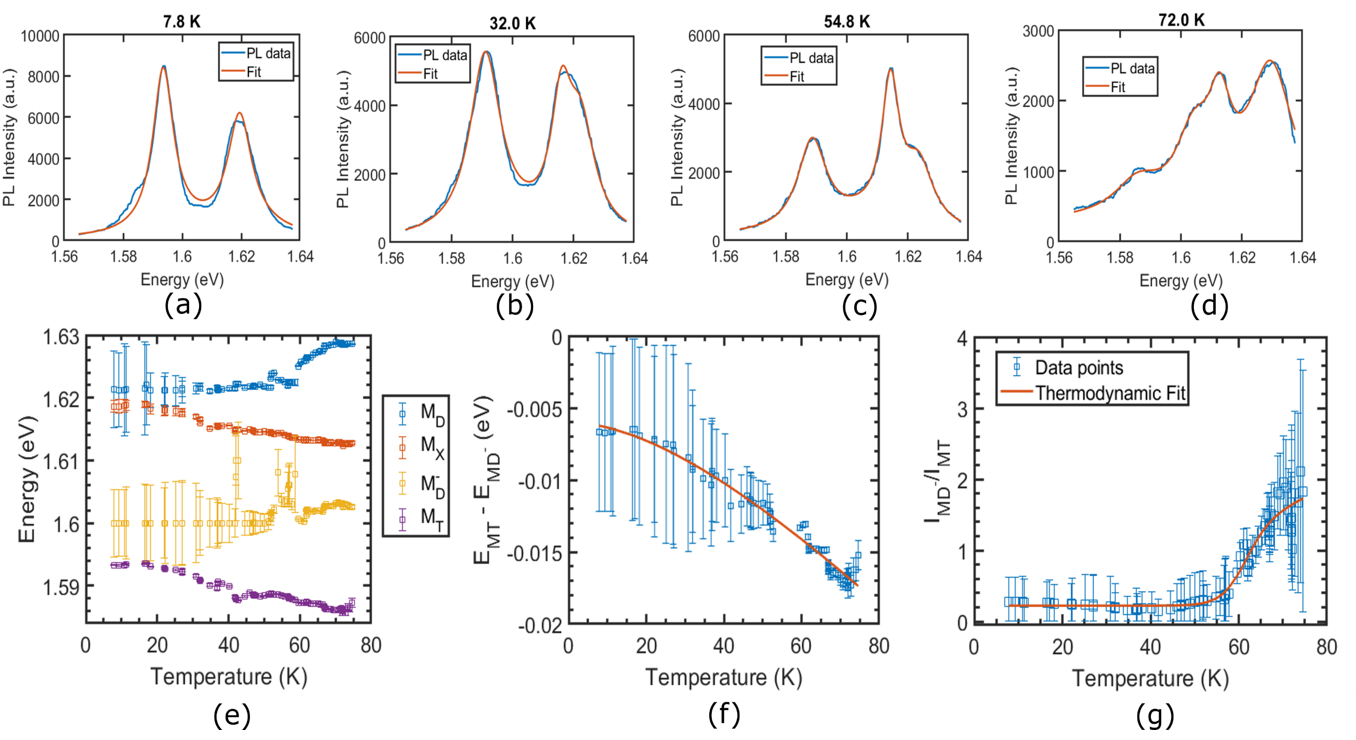}
\caption{(a)-(d) PL spectrum and lorentzian fits to the data at indicated temperatures, (e) peak energies of the four fitted species, (f) energy difference of the bright and dark trion as a function of temperature fitted with a Varshni equation (red line), (g) ratio of integrated PL intensities and their thermodynamic fit. }
\label{fig5}
\end{figure}

To establish the trionic origin of $\text{M}_{D}^{-}$, we consider how the ratio of the integrated PL intensities of the bright and dark trions, $\text{M}_\text{T}$ and $\text{M}_{D}^{-}$ evolves with temperatures. Trions being fermions are assumed to obey a Fermi-Dirac step-like dependence in their optical absorption given by, $\alpha \propto \frac{1}{1+\exp{\frac{T-T_0}{\gamma}}}$, where values of $T_0$ and $\gamma$ are determined by intervalley scattering processes among other things\cite{7}.  The integrated PL intensities of the species $A$ are then given as a function of temperature by: $I_A (T) = n_A(T)\alpha_{A}(T)$, where $n_{M_{T}} = \frac{n_0}{1+\exp{\frac{E(M_T)-E(M_D^-)}{k_B\,T}}}$ and $n_{M_{D}^-} = \frac{n_0\exp{\frac{E(M_T)-E(M_D^-)}{k_B\,T}}}{1+\exp{\frac{E(M_T)-E(M_D^-)}{k_B\,T}}}$ , $n_0$ being the total density of trions in the system. The integrated PL intensities are obtained from the spectrums by fitting them with four lorentzians as shown in Supporting figure.~\ref{fig5} (a)-(d). The energy difference between the two species is fit with a Varshni-like equation, $E(M_T)-E(M_D^-) = \Delta_0 - \frac{\alpha_1\,T^2}{T+\beta_1} + \frac{\alpha_2\,T^2}{T+\beta_2} $, where $\alpha_2$ is allowed to be negative to account for the blueshift of the dark trion energy with temperature. This fit of the difference in energy is used to fit the ratio of the PL intensities with their expected form in Supporting figure.~\ref{fig5} (f). We hence obtain the relevant values as shown in the table below.

\begin{center}
\begin{tabular}{ | m{1cm} | m{4cm}| m{4cm} | } 
  \hline
     &$T_0 (K)$ & $\gamma (K)$ \\ 
  \hline
  $\text{M}_{\text{T}}$ & $61.18 \pm 1.84$ &  $2.8417 \pm 0.89$ \\ 
  \hline
  $\text{M}_{\text{D}}$ &  $56.18 \pm 1.78$ & $2.7166 \pm 0.92$ \\ 
  \hline
\end{tabular}
\end{center}

\subsection{PL from different spots}
In the first sample, we observe the same evolution of the PL at different spots with temperature, wherever the linewidths are narrow enough for disambiguation. For spot 4, where the feature is undiscernible at 8 K, the temperature evolution still points to the population transfer. 

\begin{figure}[h]
\centering
\vspace{0.5cm}
\includegraphics[height=0.40\textwidth,width=1.00\textwidth]{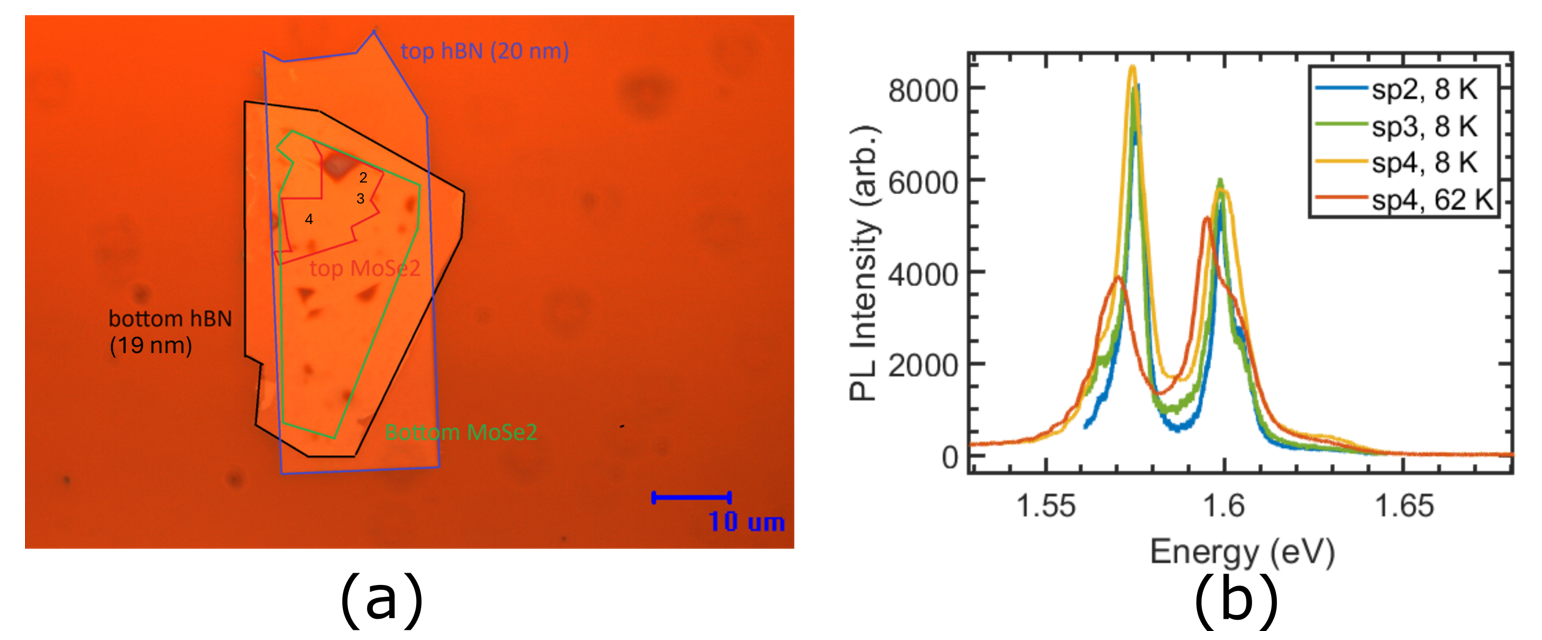}
\caption{(a) Different spot positions within the sample, (b) PL from different spots. }
\label{fig5}
\end{figure}
\newpage
\subsection{AFM data of the first sample}

\begin{figure}[h]
\centering
\vspace{0.5cm}
\includegraphics[height=0.35\textwidth,width=1.00\textwidth]{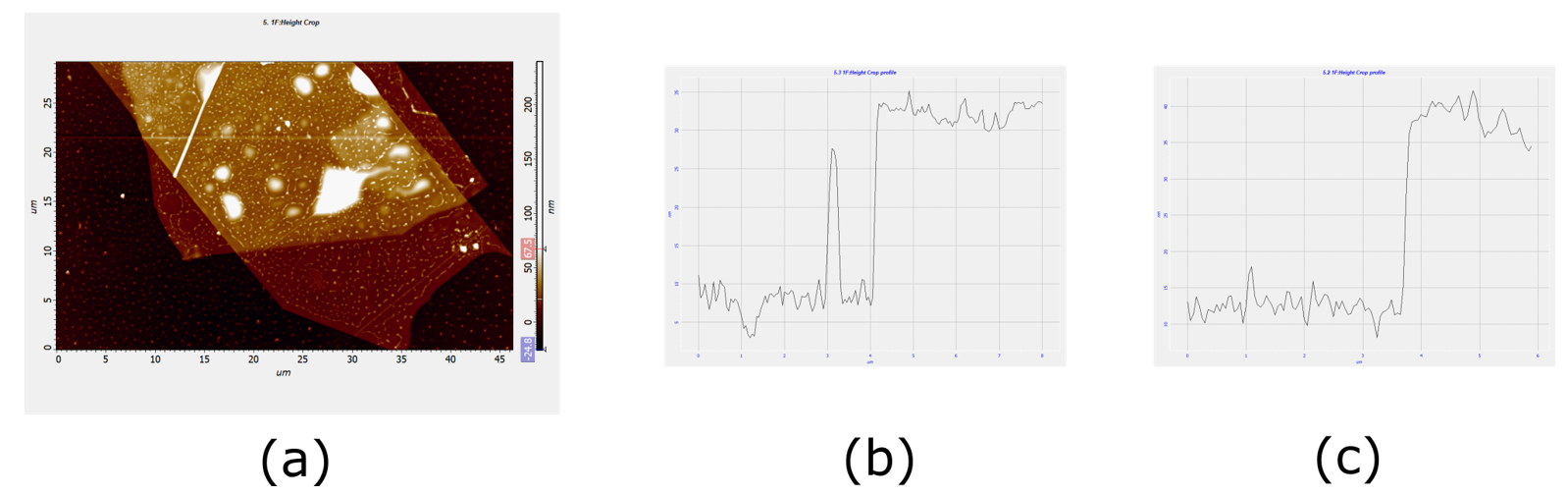}
\caption{(a) AFM of the first sample, (b) linecut of top hBN, (c) linecut of bottom hBN  }
\label{fig5}
\end{figure}

\subsection{Fits of the PL spectra under a magnetic field, 532 nm excitation}

\begin{figure}[h]
\centering
\vspace{0.5cm}
\includegraphics[height=0.40\textwidth,width=1.00\textwidth]{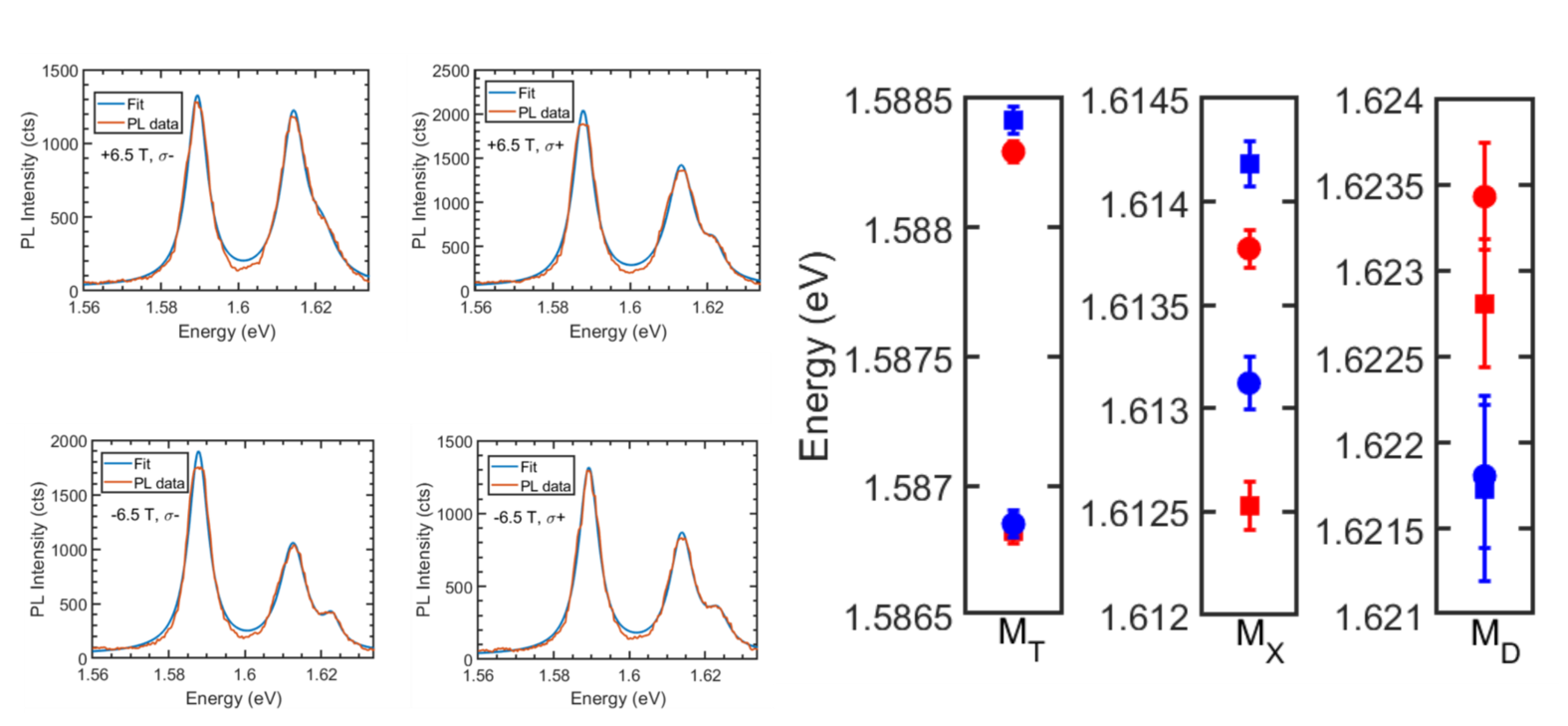}
\caption{(left) Fits of PL spectra with three lorentzians and background, (right) extracted energies and 95 percent confidence intervals of trion, bright and dark excitons, red (blue) color represents -6.5 T (+6.5 T) while squares (circles) represents the $\sigma_- (\sigma_+)$ signal.}
\label{fig5}
\end{figure}

\subsection{ChatGPT 4o transcript}
In keeping with the best practices\cite{RN135} in reporting the use of AI-enabled scientific communication we share the prompt and the response below:

\textbf{Prompt:}

This is a manuscript I have been preparing. Can you give suggestions on consistency, grammatical errors, and spelling mistakes? Please give the specific changes suggested. 

\textbf{Response:}

\begin{spverbatim}

Sure, here are specific suggested changes for your manuscript titled "Diffusion of brightened dark excitons in a high-angle incommensurate Moiré homobilayer":


Abstract:

Current: "twistronics, especially in low-angle twisted bilayers of transition metal dichalocogenides."
Suggested: "twistronics, especially in low-angle twisted bilayers of transition metal dichalcogenides."

Current: "While a lot remains to be known and understood about these heterostructures..."
Suggested: "While much remains to be known and understood about these heterostructures..."

Introduction:

Current: "These systems have been explored in a recent work22 over a range of twist angles."
Suggested: "These systems have been explored in a recent work [22] over a range of twist angles."
- Ensure consistent use of "twist angle" and "high-angle."

Results and Discussion:

Current: "We next shift our attention to the diffusion of the excitonic and trionic complexes..."
Suggested: "Next, we shift our attention to the diffusion of the excitonic and trionic complexes..."

Current: "Upon excitation with a power of 10µW in a spot of radius ∼ 1µm..."
Suggested: "Upon excitation with a power of 10 µW in a spot of radius ∼1 µm..."

Methods:

Current: "The monolayers and hBN (high-pressure anvil growth) were mechanically exfoliated..."
Suggested: "The monolayers and hBN (high-pressure anvil growth) were mechanically exfoliated from high-quality bulk samples obtained from 2D Semiconductors."

Current: "an achromatic lens system was used to image the emission spot..."
Suggested: "an achromatic lens system was used to image the emission spot on the CCD camera..."


Grammar and Spelling Corrections:

Current: "The diffusion lengths points us toward the presence of a more diffusive excitonic species..."
Suggested: "The diffusion lengths point us toward the presence of a more diffusive excitonic species..."

Current: "the spectral proximity of the bright and dark excitons causes the photoluminescence from one of the species to affect the apparent diffusion length..."
Suggested: "the spectral proximity of the bright and dark excitons causes the photoluminescence from one of the species to affect the apparent diffusion length..."

Supplementary Information:

- **Excitation Intensity Dependence of Diffusion Lengths**:
  - **Current**: "the uncorrected (MX and MD are spectrally mixed and affect each others’ values of diffusion lengths)..."
    - **Suggested**: "the uncorrected (MX and MD are spectrally mixed and affect each other's values of diffusion lengths)..."
- **Disentangling Contributions of Bright and Dark Excitons**:
  - **Current**: "achromatic lens used to image the emission spot at the entrance slit of the spectrometer-CCD setup..."
    - **Suggested**: "The achromatic lens used to image the emission spot at the entrance slit of the spectrometer-CCD setup..."
- **Temperature Dependent Measurements**:
  - **Current**: "We see that between 6K and 52 K..."
    - **Suggested**: "We see that between 6 K and 52 K..."
  - **Current**: "lorentzian functions results in a large error..."
    - **Suggested**: "Lorentzian functions result in a large error..."

### Consistency Checks:
- Ensure all figure citations are consistent, e.g., "Figure 2(a)" should be consistently cited as "Figure 2a."
- Verify consistent use of symbols and notation, e.g., "µm" instead of "microns" and "nm" consistently throughout.

By making these specific changes, you can improve the consistency, grammar, and clarity of your manuscript. If you need further assistance with any section, feel free to ask!

\end{spverbatim}

\bibliography{biblio}